\documentclass[prl,amsmath,amssymb,twocolumn]{revtex4}
 \usepackage[english]{babel} 
 \usepackage{latexsym} 
 \usepackage{amsmath} 
 \usepackage{amsfonts} 
 \usepackage{amssymb} 
 \usepackage{mathtools}
 \usepackage{textcomp} 
 \usepackage{hyperref}
 \usepackage{graphicx}  
 \usepackage{enumerate} 
 \usepackage{array} 
 \usepackage[latin3]{inputenc} 
 \usepackage{graphics} 
 \usepackage{color} 

 \newcommand{\ar}{\arrowvert} 
  
 \newcommand{\be}{\begin{equation}} 
 \newcommand{\ee}{\end{equation}} 
 \newcommand{\ba}{\begin{eqnarray}} 
 \newcommand{\ea}{\end{eqnarray}}

\newcommand{\Imag}{\mathop{\mathrm{Im}}}

\begin{document} 
%opening 
%%%%%%%%%%%%%%%%%%%%%%%%%%%%%%%%%%%%%%%%%%%%%%%%%%%%%%%%%%%%%%%%%% 
\title{Possible new resonance from $W_L W_L$-$hh$ interchannel coupling} 
\author{Rafael L. Delgado, Antonio Dobado and Felipe J. Llanes-Estrada} 
\affiliation{ 
%Departamento de 
F\'{\i}sica Te\'orica I, Univ. Complutense, 
Parque de las Ciencias 1,  
28040 Madrid, Spain.} 
%%%%%%%%%%%%%%%%%%%%%%%%%%%%%%%%%%%%%%%%%%%%%%%%%%%%%%%%%%%%%%%%%% 
\begin{abstract}  
We propose and theoretically study a possible new resonance caused by strong coupling between
the Higgs-Higgs and the $W_L W_L$ ($Z_L Z_L$) scattering channels, without regard to the intensity of the
elastic interaction in either channel at low energy (that could be weak as in the Standard Model). We expose this channel-coupling resonance from unitarity and dispersion relations encoded in the Inverse Amplitude Method, applied to the Electroweak Chiral Lagrangian with a scalar Higgs.
\end{abstract}

\maketitle 

%%%%%%%%%%%%%%%%%%%%%%%%%%%%%%%%%%%%%%%%%%%%%%%%%%%%%%%%%%%%%%%%%%%%%%%%%%%%%%%%%%%%%%%%%%%%%%%%%%%%%%%%%%%%%%

\begin{figure}
\vspace{-0.4cm}
\includegraphics*[width=0.8\columnwidth]{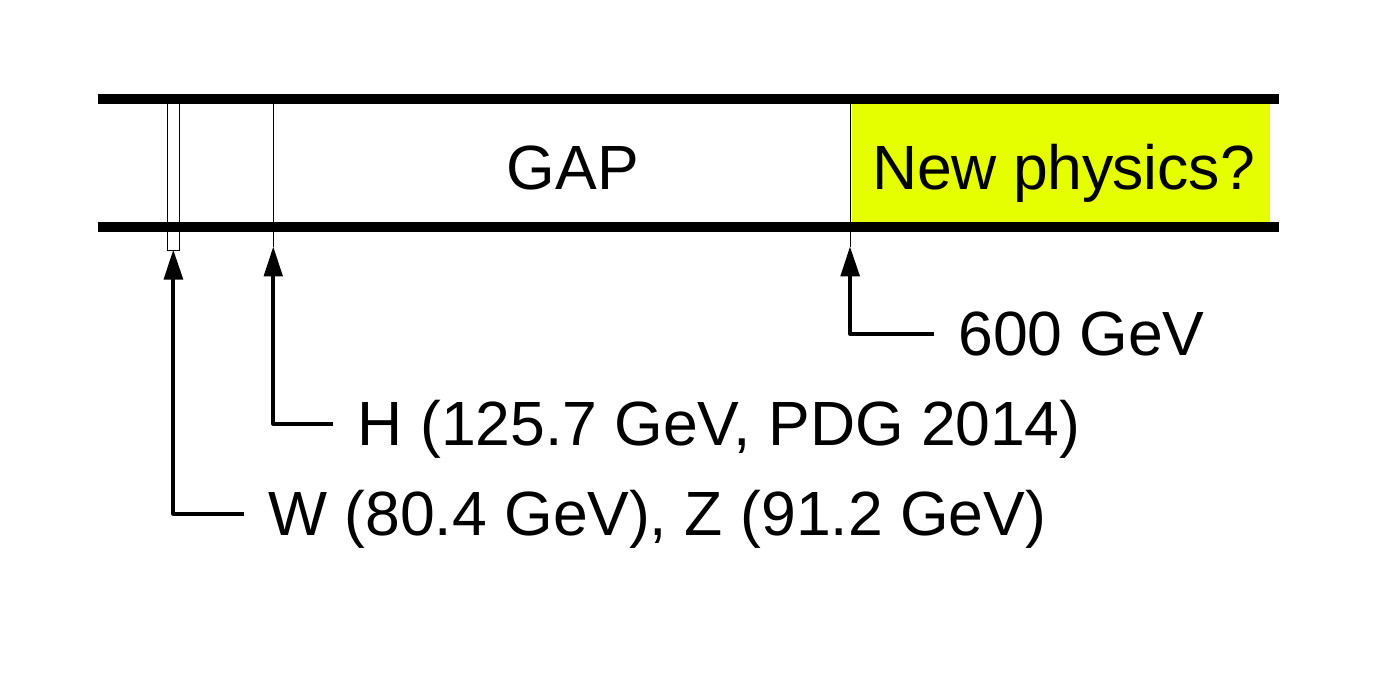}
\vspace{-0.6cm}
\caption{\label{Gap2014}%
Electroweak-symmetry breaking sector of the Standard Model after the LHC run I: there are four ``low-energy'' bosons and any new physics is split by a mass gap.}
\end{figure}

The LHC experiments CMS and ATLAS~\cite{CMSATLAS} have seemingly found what looks like a Higgs boson (mainly an excess of four-lepton events and two-photon events at 125 GeV suggestive of scalar quantum numbers). This  finding has been widely discussed, but less recognized is the equally interesting fact that no new physics beyond the Standard Model (SM) appears up to energies of 600-700 GeV for generic 
searches~\cite{Aad:2014aqa}, as shown in Fig.~\ref{Gap2014}. Therefore the unknown system giving rise to the  electroweak symmetry breaking (the Symmetry Breaking Sector (SBS)) from $SU(2)_L \times U(1)_Y$ to $U(1)_{em}$ should contain four low-mass states: the three would-be Goldstone bosons $\omega^a$ ($a=1,2,3$) responsible
for the $W^\pm$ and $Z$ masses and the recently discovered particle $h$. Because this Higgs-like boson turns out to be light and the spectrum is gapped up to the scale of any new physics, it is natural to think of it also as an approximate (composite) Goldstone boson (GB) itself~\cite{kaplan}. For instance, as one of GB  corresponding to the  spontaneous symmetry breaking from a group $G$ to a group $H$ with $\dim(G)-\dim(H)=4$. This is for example the case of the so-called MCHM (Minimal Composite Higgs Model~\cite{Agashe:2004rs}, with $G=SO(5)$ and $H=SO(4)$). Another exciting possibility is that the Higgs is the dilaton~\cite{Grinstein}  (the Goldstone boson associated with spontaneous breaking of the scale symmetry of the SBS).

Assuming the approximate, well established $SU(2)_{L+R}$ custodial symmetry,  the low-energy GB dynamics can be properly described by a $SU(2)_L \times U(1)_Y$ gauged non-linear effective Lagrangian~\cite{Alonso:2012px,Buchalla:2013rka}, which is an extension of the former Higgsless Electroweak Chiral Lagrangian~\cite{EWCL}. Thus the three $\omega^a$ GB fields parametrize the coset $SU(2)_L \times SU(2)_R/SU(2)_{L+R}$ and the Higgs-like boson $h$ is a custodial isospin singlet.

In this work we are concerned with the $W_LW_L$, $Z_LZ_L$ and $hh$ scattering. This is because thanks to the Equivalence Theorem~\cite{ET}, we can get information about the unknown SBS of the SM by studing the GB dynamics, 
whose amplitudes approximate well those of the longitudinal vector bosons $W_L$ and $Z_L$ ($W_L$ for short in the following) of the SM for energies well above the $W$ mass $(E\gg M_W)$. In this regime we can also neglect the $h$ boson mass since $M_h= 125$ GeV $\sim M_W$. Then the relevant Lagrangian for $W_LW_L$  and $hh$ scattering is~\cite{nos}:
\ba \label{Lagrangian}
{\cal L} & = & \left(\! \! 1\! +\! 2 a \frac{h}{v} +b \frac{h^2}{v^2}\right)
\frac{\partial_\mu \omega^a \partial^\mu \omega^b}{2}
\left(\!\! \delta^{ab}\!+\! \frac{\omega^a\omega^b}{v^2}\!\! \right)   
\nonumber  \\ \nonumber
 & + &  \frac{4 a_4}{v^4}\left( \partial_\mu \omega^a\partial_\nu \omega^a \right)^2 
+ \frac{4 a_5}{v^4}\left( \partial_\mu \omega^a \partial^\mu \omega^a\right)^2  
\\ \nonumber
 & + & \frac{2 d}{v^4} \partial_\mu  h \partial^\mu h\partial_\nu \omega^a  \partial^\nu\omega^a
+\frac{2 e}{v^4}\left( \partial_\mu h \partial^\mu \omega^a\right)^2
\\
 & + & \frac{1}{2}\partial_\mu h \partial^\mu h +\frac{g}{v^4} (\partial_\mu h \partial^\mu h)^2  
\ea which should be valid for $M_W, M_h\ll E\ll 4 \pi v \simeq 3\,{\rm TeV}$. Thus we have set $M_h=M_W=M_Z\simeq 0$. Different SBS dynamics  can be modeled by a proper tuning of the parameters $a$, $b$, $a_4$, $a_5$, $d$, $e$ and $g$. The last five of them must be renormalized to some scale $\mu$ to absorb the one-loop divergencies coming from the lowest order (LO), i.e. the first term in the Lagrangian, in a similar way as in Chiral Perturbation Theory~\cite{ChPT}. In the SM  $a^2=b=1$ and   the rest of the  tree level parameters vanish ($a_4=a_5=d=e=g=0$). As is well known in this particular case we get a linear theory which is renormalizable in the standard way and weakly interacting for light $h$.

From the Lagrangian density in Eq.~(\ref{Lagrangian}), the LO partial waves with $I=J=0$ ($I$ being the custodial isospin) can be easily  computed and one finds: 
\ba
\label{treeamps}
A_0(\omega\omega\to \omega\omega)&=&  \frac{s}{16 \pi v^2} (1-a^2)  \\  \nonumber
T_0(hh\to hh)&=& 0 \\ \nonumber
M_0(\omega\omega\to h h) &=&\frac{ \sqrt{3} s}{32 \pi v^2} (a^2-b) \ .
\ea
From these low-energy theorems we can expect strong $W_LW_L$ elastic scattering whenever $a^2\neq 1$ (as Eq.~(\ref{treeamps}) grows with $s=E_{\rm cm}^2$). For $b\neq a^2$ we have strong mixing between the $W_LW_L$ and $hh$ channels~\cite{Contino, Delgado:2013loa} even in the case $a=1$ (no LO  contribution to $W_LW_L$ elastic scattering). As we will show bellow, this strong mixing gives rise to a new resonance in the $I=J=0$ channel for an important region of the available $(a,b)$ parameter space after taking into account the experimental information coming from the LHC. At next to leading order (NLO) all three amplitudes in Eq.~(\ref{treeamps}) acquire one-loop contributions of order $O(s^2)$.  The elastic ones $A_1$ and $T_1$ are accompanied by logarithmic left and right cuts (LC and RC respectively) in the $s$-plane, entailing an imaginary part for physical energy corresponding to $s$  just above the RC. Those amplitudes have already been reported in recent literature~\cite{nos, Espriu}. The divergences appearing for massless $W$ and $h$ can be absorbed by renormalization of the $a_4$, $a_5$, $d$, $e$ and $g$ parameters but no $a$ or $b$ renormalization is needed in this case. 

In this work we want to focus on the very interesting phenomenon of the strong mixing appearing whenever $a^2 \ne b$. In order to emphasize this point we will concentrate first in the particular case where $a=1$ (no direct $W_LW_L$ LO elastic scattering) and the rest of the  parameters except $b$ vanish~\cite{Khemchandani:2011et}. As the renormalized parameters depend on the renormalization scale $\mu$, that particularly simple choice requires to set this scale to some given value that now plays the role of an ultraviolet cutoff. For definiteness we will take $\mu= 4 \pi v \simeq 3\,{\rm TeV}$ which is roughly the limit of applicability of our effective theory. The relevant NLO $I=J=0$ partial waves read~\cite{nos}:
\ba \label{pertA}
A_1(\omega\omega\to \omega\omega) = \frac{s^2(1-b)^2}{256\pi^3 v^4} \times
\\ \nonumber
\left[ \frac{17}{9}-\frac{1}{6} \log \left( \frac{s}{\mu^2}\right)
-\frac{3}{4} \log \left( \frac{-s}{\mu^2}\right) \right] \\ \label{pertT}
T_1(hh\to hh) = \frac{s^2(1-b)^2}{32\pi^3 v^4} \times \\ \nonumber
\left[ \frac{1}{3}-\frac{1}{16} \log \left( \frac{s}{\mu^2}\right)
-\frac{3}{32} \log \left( \frac{-s}{\mu^2}\right) \right] \\
\label{pertM}
M_1(\omega\omega\to h h)= 
\frac{\sqrt{3}s^2(1-b)^2}{9216\pi^3 v^4}
\left[ \frac{1}{2}-\log \left( \frac{s}{\mu^2}\right)\right] \ .
\ea

%%%%%%%%%%%%%%%%%%%%%%%%%%%%%

%%%%%%%%%%%%%%%%%%%
\begin{figure}
\null\hfill%
\includegraphics[width=0.6\columnwidth]{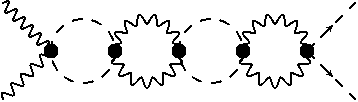}%
\hfill\null\vspace{-0.3cm}
\caption{\label{fig:FeynmanDiagram}%
Typical Feynman diagram mixing the $\omega\omega$  (wiggled lines) and the Higgs-Higgs (dashed lines) channels.\vspace{-0.5cm}}
\end{figure}

%%%%%%%%%%%%%%%%%%%%

These partial waves have adequate analytical properties featuring a LC and also a RC  just under the physical region  $s= E_{\rm cm}^2+i\epsilon$. However, unitarity is satisfied only perturbatively, with $\Imag A_1 = \Imag T_1 = \lvert M_0\rvert^2$ on the RC. Notice also that the cross-channel amplitude $M_1$ has only a LC and is thus purely real ($\Imag (M_0+M_1)=0$) on the RC. As the amplitudes grow with $s$ they will eventually violate the unitarity bound (for example $\Imag A\le 1$). Grouping the two coupled channels in matrix form,
\ba \label{perttotal}
F&=&F_0+F_1+\dots  \\ \nonumber
&=&\left( \begin{tabular}{cc} $0$ &$M_0$ \\ $M_0$ & $0$ \end{tabular}\right)\ \ %\nonumber
+\left( \begin{tabular}{cc} $A_1$ &$M_1$ \\ $M_1$ & $T_1$ \end{tabular}\right)
+\dots
\ea
the perturbative unitarity relation satisfied is $\Imag F_1 = F_0^\dagger F_0$; but exact unitarity requires $\Imag F = F^\dagger F$ instead. However there is a very well known method, based on dispersion relations, called the Inverse Amplitude Method (IAM)~\cite{IAM}, that allows to fully unitarize the perturbative partial waves,  even in the coupled channel case~\cite{Pelaez:1999bp}.  The resulting amplitudes are given by $F^{\rm IAM}\coloneqq F_0(F_0-F_1)^{-1}F_0$. The IAM amplitudes still have the analytical properties found above but now they fulfill exact unitarity. In addition, 
the determinant of  $F_0-F_1$ appearing in the denominators allows for the possibility of having  poles in the second  Riemann sheet for some regions of the parameter space. When they are close enough to the physical region, those poles have the  natural interpretation of dynamical resonances. The IAM amplitudes have been extensively used to fit meson-meson scattering data~\cite{JRAGM}. The method has also been applied to $W_LW_L$ elastic scattering~\cite{Espriu2} where resonances were found in different channels in terms of the $a_4$ and $a_5$ parameters. In this work we are rather interested in the pure coupled channel resonances appearing even for $a=1$. The IAM amplitudes are given in this case by:
\be\label{IAM}
F^{\rm IAM} = \frac{M_0^2}{(\! M_0\!-\!M_1\!)^2\!-\!A_1T_1}\!
\left(\!
\begin{tabular}{cc}
$A_1$     & $M_0-M_1$   \\
$M_0-M_1$ & $T_1$  
\end{tabular}\!
\right)
\ee
whose perturbative expansion coincides with Eq.~(\ref{perttotal}) at $O(s^2)$ but that satisfies exact unitarity, $\Imag F^{\rm IAM}=(F^{\rm IAM})^\dagger F^{\rm IAM}$ as is easily checked. The IAM resums the imaginary parts of diagrams like that in Fig.~\ref{fig:FeynmanDiagram}

\begin{figure}
\includegraphics*[width=0.49\columnwidth]{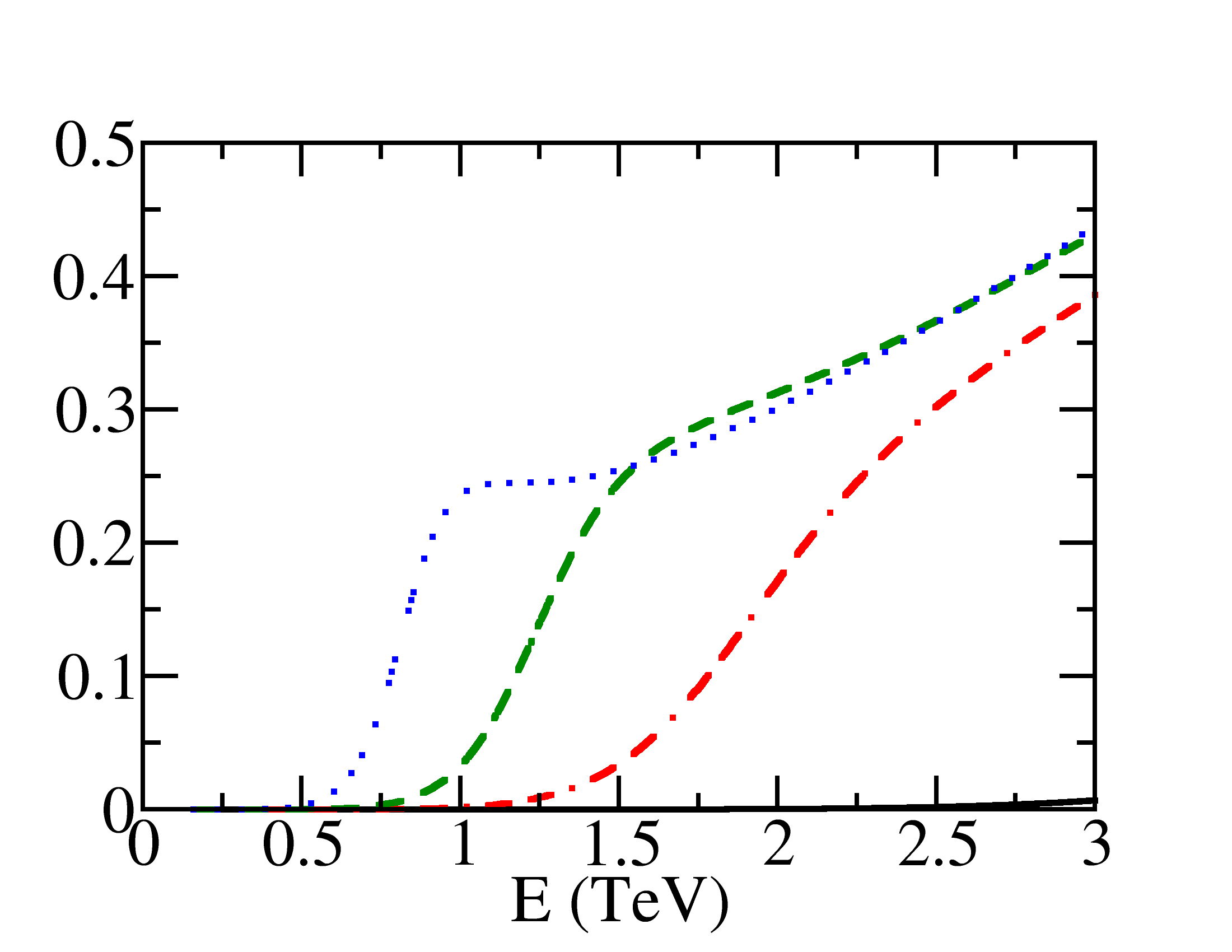}\vspace{-0.1cm}
\includegraphics*[width=0.49\columnwidth]{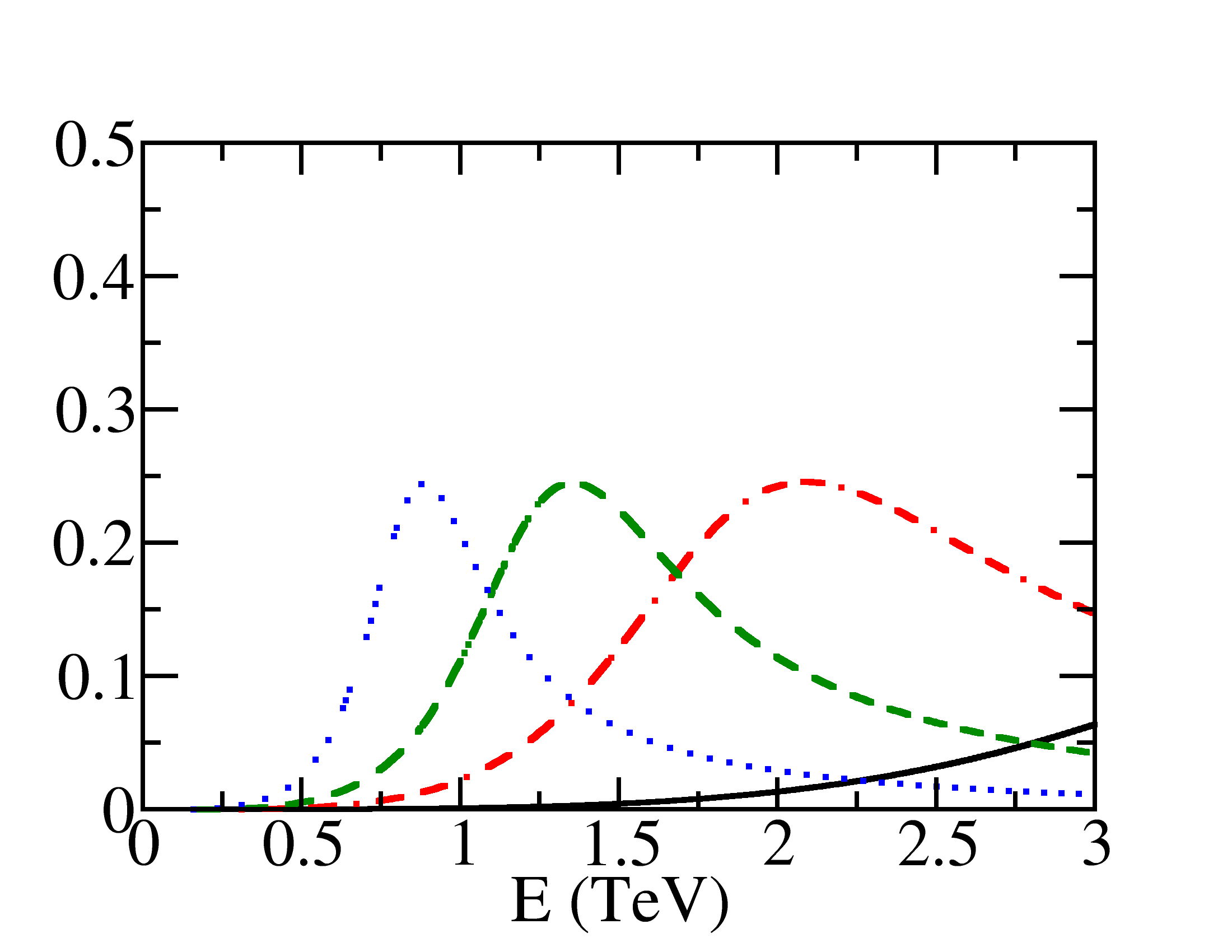}\vspace{-0.1cm}
\caption{\label{fig:realaxis}%
Left graph: elastic $\lvert A^{\rm IAM}\rvert^2$. Right graph: cross-channel $\lvert M^{\rm IAM}\rvert^2$. Various values of $b$ shown are $b=1.1$ (lowest, solid black line), $b=1.5$ (dot-dashed, red online), $b=2$ (dashed, green online) and $b=3$ (dotted, blue online). The inelastic amplitude $M^{\rm IAM}$ shows the resonance with most clarity: it becomes narrower and less massive for larger $b$.
\vspace{-0.5cm}
}
\end{figure}

\begin{figure}
\includegraphics[width=0.8\columnwidth]{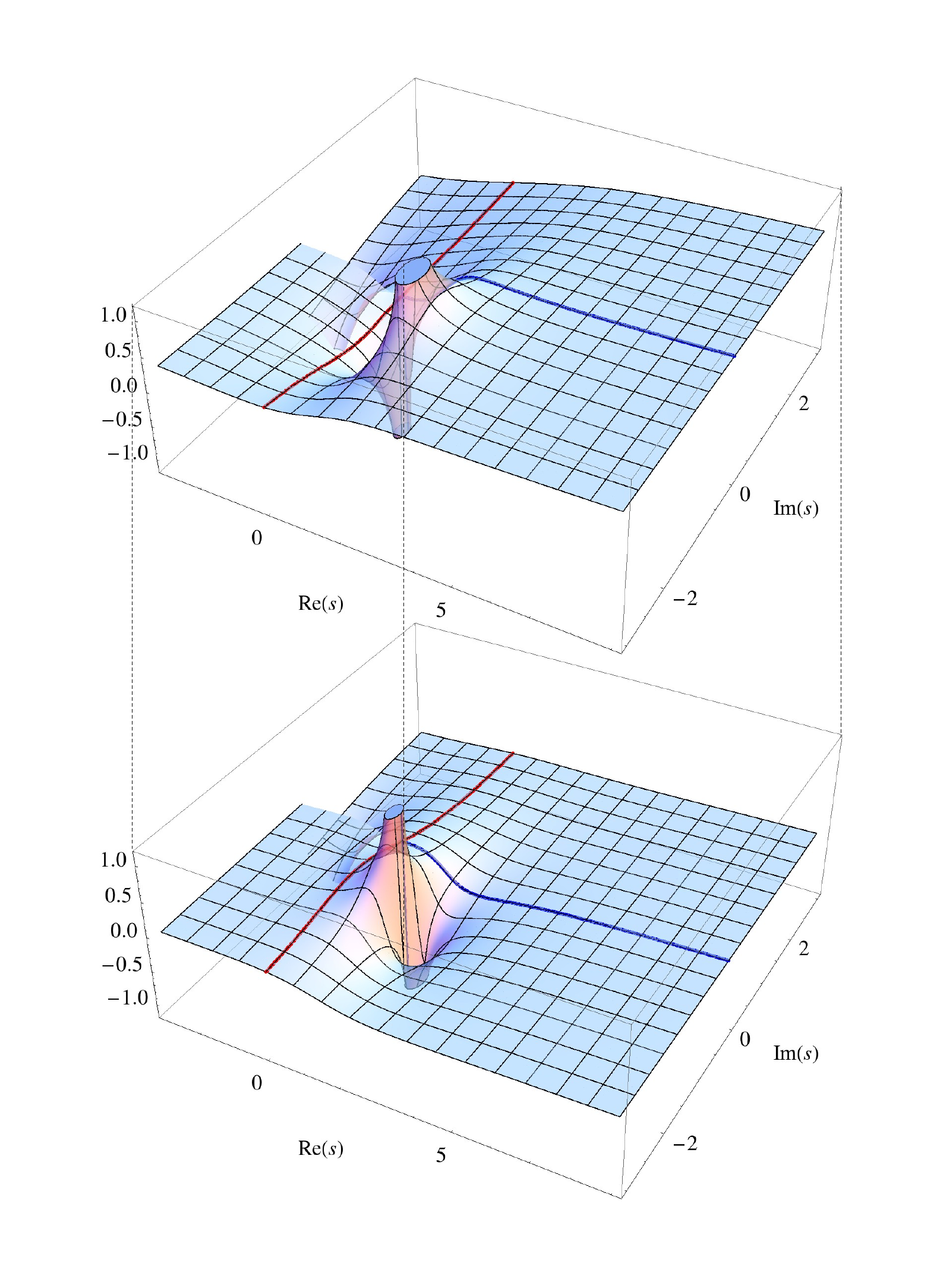}
\vspace{-0.9cm}
\caption{\label{fig:pole}%
Imaginary part of the unitarized partial waves in the second Riemann sheet in terms of the Mandelstam variable $s$ for $b=2$. Top: the elastic amplitude $A^{\rm IAM} $. Bottom: the inelastic amplitude $M^{\rm IAM}$. These amplitudes are different but still they show a pole at the same point of the second Riemann sheet that could be understood as a new resonance. 
\vspace{-0.5cm}}
\end{figure}

In Fig.~\ref{fig:realaxis} we show the square moduli of two distinct matrix elements of $F^{\rm IAM}$ ($A^{\rm IAM}$ and $M^{\rm IAM}$). For $b$ nearly 1 the amplitude vanishes in the LHC region of interest (with $W_L W_L\to W_L W_L$ reducing to the small SM amplitude).  For larger $b$ there is a resonant structure most clearly seen in the inelastic process $\omega\omega\to hh$, but that leaves a trace also in the elastic amplitudes that grow almost vertically for large $b$ and even peak slightly up to the resonance mass, growing more calmly afterwards. 

In order to explore in more detail the peaks in Fig.~\ref{fig:realaxis}, we have analytically extended  the complex $\log(-s/\mu^2)$ of Eqs.(\ref{pertA}) to (\ref{pertM}) to the second Riemann sheet, and indeed we found a  pole   (see Fig.~\ref{fig:pole}) at the same point in all the channels. This pole is naturally interpreted as a dynamical resonance whenever it is close enough to the physical, real $s$. Then the position of the pole $s_0$ is related to the parameters of the resonance (mass $M$ and width $\Gamma$) as $s_0=M^2-i \Gamma M$ which for small $\Gamma/M$ means $\sqrt{s_0}\simeq M- i \Gamma/2$. We follow these variables with $b$ in the complex $s$ plane, 
%is displayed in Fig.~\ref{fig:polemotion}. For this, we 
numerically finding the zeroes of the denominator of Eq.~(\ref{IAM}) in the second Riemann sheet %We show values for $b>1$. 
(that denominator is an even function of $b-1$, so there is an approximate symmetry between $b>1$ and $b<1$).
%the plot for $b<1$ falls on top of the shown trajectory. 
The pole escapes to infinity as $b$ approaches $1$, then returns along nearly the same trajectory as $b$ increases beyond $1$.
\begin{figure}
\includegraphics[width=0.8\columnwidth]{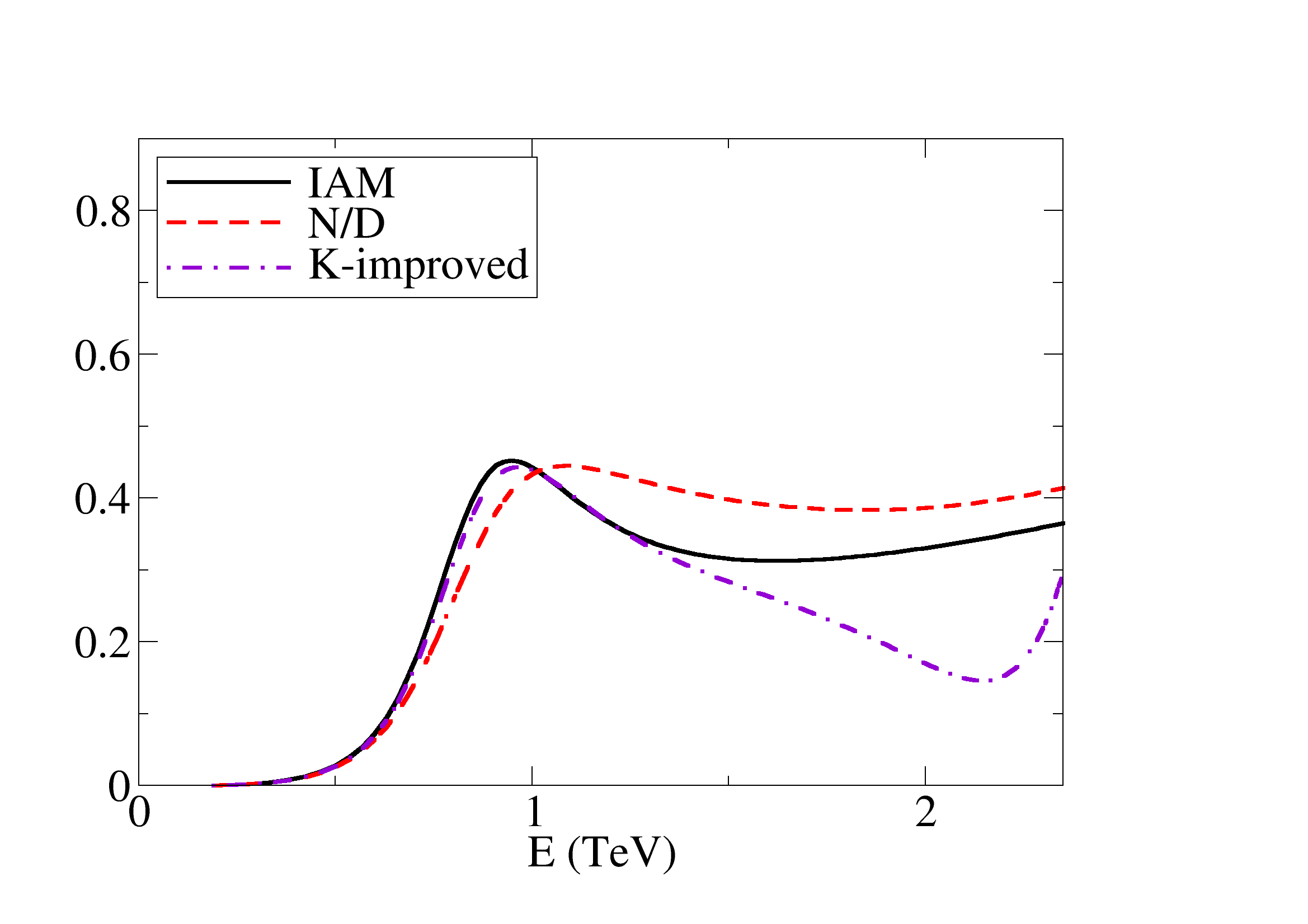}\vspace{-0.6cm}
\caption{\label{fig:compare}%
Three unitarisation methods closely agree on the existence of a coupled-channel resonance (we show ${\rm Im} A = \ar A\ar^2 + \ar M\ar^2$ for $a\to 1$, $b=3$). The $I=J=0$ IAM mass ($0.95\,{\rm TeV}$) is within 2\% of the Improved-K and within 10\% of the N/D methods.}
\vspace{-0.5cm}
\end{figure}

%\begin{figure}
%\includegraphics[width=\columnwidth]{FIGS_DIR/PoleMotion2ndSheet_b_eta.png}\vspace{-0.8cm}
%\caption{\label{fig:polemotion}
%Dependence of the resonance position on $b$ with $a^2=1$ fixed (upper curve) 
%and for $a=\sqrt{1-\xi}$ and $b=1-2\xi$ with $\xi=v^2/f^2$ as in the MCHM (lower curve, blue online). \vspace{-0.5cm}}
%\end{figure}

\begin{figure}
\includegraphics[width=0.8\columnwidth]{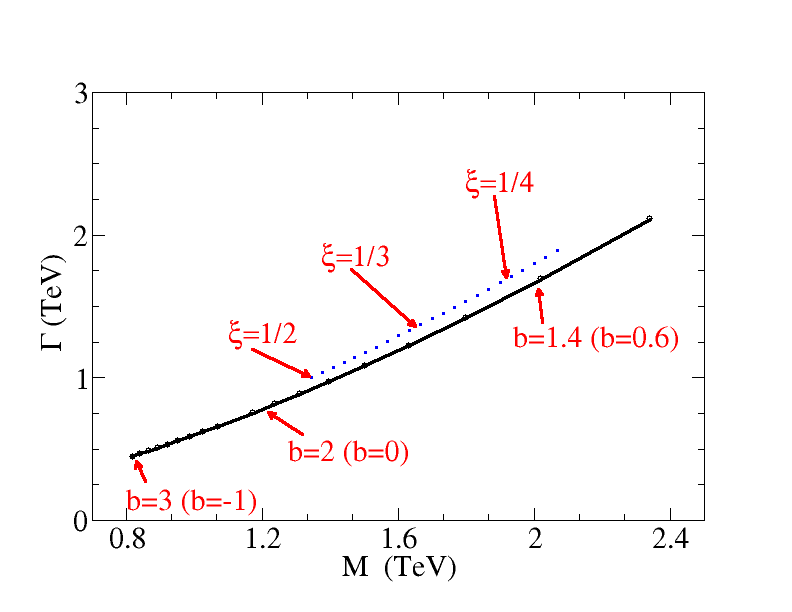}\vspace{-0.6cm}
\caption{\label{fig:polemotion2}%
Dependence of resonant mass and width on $b$,  with $a^2=1$ fixed (lower curve) and for $a=\sqrt{1-\xi}$, $b=1-2\xi$ with $\xi=v^2/f^2$ as in the MCHM (upper curve, blue online).\vspace{-0.5cm}}
\end{figure}

%we show the  mass and  the width of the resonance associated to the pole. 
%For $b$ larger than about 3 (or smaller than -1), the resonance is light and moderately narrow, and it could have already been detected at the LHC.
Currently there is no constraint on the $b$ parameter but, if $a$ is set to its SM value of 1, and in the absence of NLO couplings, we can provide the bound $b\in (-1,3)$ (around the SM $b=1$) because otherwise the resonance moves below $700\,{\rm GeV}$ where it would have already  been seen by ATLAS and CMS. These exclusion limits on $b$ are of course uncertain by $M_W/E\simeq 15\%$ from our using the Equivalence Theorem ($\omega^a \simeq W_L^a$) and, consistently, the massless Higgs approximation $M_h\ll E$.

Another potential source of uncertainty is the IAM unitarization employed. In the accompanying article we also address the N/D method and the Improved-K matrix. Because the scalar resonance follows from unitarity and analiticity (causality) in the presence of strong channel coupling, all three methods (that encode those properties and agree with perturbation theory at low energy) find it at a very similar position (see Fig.~\ref{fig:compare}).

\begin{figure}
\includegraphics[width=0.7\columnwidth]{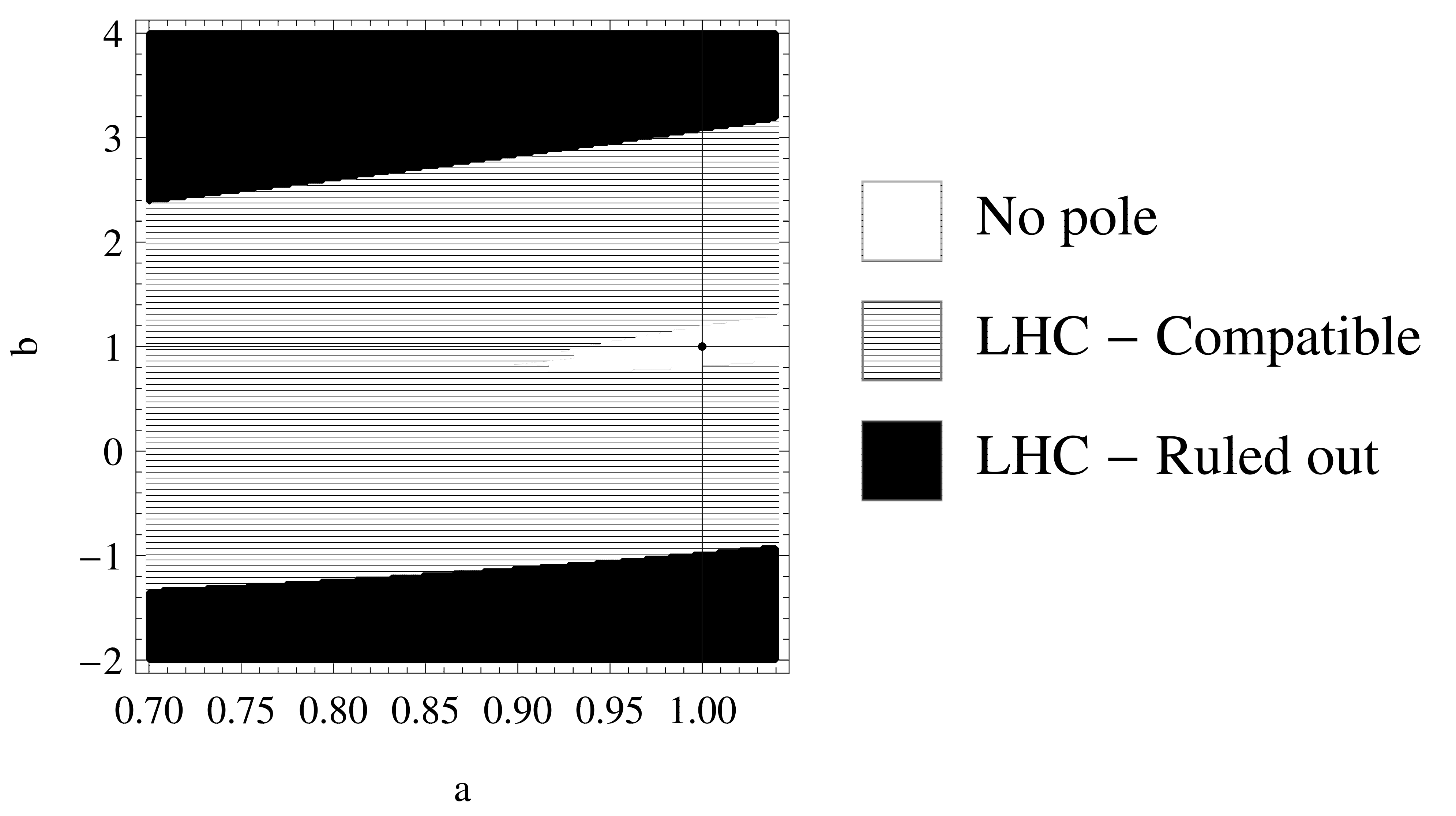}
\vspace{-0.4cm}
\caption{\label{fig:map_ab}%
Parameter $a-b$ plane showing whether the resonance mass is beyond the $3\,{\rm TeV}$ reach of our approach (white), below $700\,{\rm GeV}$ and disfavored by LHC data (black) or between those two values, where we can trust theory and data still allows (striped). The SM value $a=b=1$ is also marked.\vspace{-.5cm}}%end caption
\end{figure}

Once we have identified the new coupled-channel resonance by switching off the LO elastic $W_L W_L$ channel by setting $a=1$, we can consider now the more general case of arbitrary $a$.  From the LHC data we know this parameter must belong to the 2$\sigma$ interval $(0.88, 1.15)$ (ATLAS) and $(0.96, 1.34)$ (CMS) \cite{Datos}, while $b$ is much more  unconstrained. As commented in page~\pageref{Gap2014}, the simplest composite model where the three $\omega^a$ and $h$ show up as composite (pseudo) GB is the so called MCHM featuring the symmetry breaking pattern $SO(5)$ to $SO(4)$~\cite{Agashe:2004rs}. In this model the $a$ and $b$ parameters are given by $a=\sqrt{1-\xi}$ and $b=1-2\xi$, where $\xi= v^2/f^2$ and $f$ is a new symmetry-breaking, higher scale. 
The relevant IAM partial waves in the general case $a\ne 1$ may be retrieved from~\cite{nos,nos2}. In  Fig.~\ref{fig:polemotion2} we show the  mass and width in terms of $\xi $ for the MCHM model. Our new resonance appears for an important range of the allowed $\xi$ parameter range ($0 < \xi < 0.5$). Even for the region where the pole is too far away from the real axis to be considered a resonance, it will produce a huge  increment of the cross section for $W_L W_L$ and $hh$ production that could be probed at the LHC. This conclusion probably applies to other composite models beyond the MCHM too. The figure also shows the pure coupled-channel case $a=1\neq b^2$ described in page~\pageref{fig:polemotion2}.
For finite $a-1$ the resonance receives strength from both elastic and coupled-channel scattering, but we have shown~\cite{Delgado:2013loa,nos2} that the $\sigma$-like structure from elastic dynamics alone is much broader; for finite $b-a^2$ the resonance, as shown in figure~\ref{fig:polemotion2}, is significantly narrower and lighter due to the coupled-channel dynamics.
Figure~\ref{fig:map_ab} shows the $a-b$ parameter plane shading in light gray the region where the resonance in the $2^{\rm nd}$ Riemann sheet is between $700\,{\rm GeV}$ and $3\,{\rm TeV}$.

%%%%%%%%%%%%%%%%%%%%%%%%%%%%%%%%%%%%%%%%%%%%%%%%%%%%%%%%%%%%%%%%%%%%%%%%%%%%%
To conclude, we made the case for an interesting potential phenomenon to be sought at the LHC run II and beyond; a new resonance in the $W_L W_L-hh$ coupled channels, caused by the channel-mixing interaction even when direct elastic interactions in both channels are weak~\footnote{
Outside particle physics one can find strongly coupled channels with small elastic interaction. For example, ${\rm C}_2+{\rm O}_2\to {\rm C}_2+{\rm O}_2$ or ${\rm CO}+{\rm CO}\to {\rm CO}+{\rm CO}$ elastic scattering is negligible against the channel-coupling ${\rm C}_2+{\rm O}_2\to 2 {\rm CO}$, a strong exothermic oxidation reaction, freeing almost $11\,{\rm eV}$, driven by the large phase space. What is perhaps distinctive in our mechanism is that the \emph{channel-coupling probability} is large, with no phase space advantage (all particles being approximately massless).
}%end footnote
. %end main paragraph
We do not have a strong reason to \emph{predict} this resonance, rather observe that it features in the largest part of parameter space of the effective Lagrangian with the known particle content, that supports strong channel coupling. The alternative, weakly coupled resonances that do not saturate unitarity, implies parameters fine-tuned to be very close to $a=b=1$, those of the Standard Model (that also remains a viable theory with current data). 

\emph{%
%A.D. thanks the CERN TH-Unit for its hospitality during the first part of this work and J. R.
We thank J.R. Pel\'aez for discussion. ADG and RLD thank, respectively, the CERN TH-unit and the HEP group at the University of Southampton (NEXT institute) for their hospitality. Supported by the Spanish grants UCM:910309,
MINECO:FPA2014-53375-C2-1-P, MINECO:FPA2011-27853-C02-01 and BES-2012-056054 (RLD).}
\vspace{-0.5cm}

%%%%%%%%%%%%%%%%%%%%%%%%%%%%%%%%%%%%%%%%%%%% 

\end{document}